\documentstyle[12pt]{article}
\begin{document}

\def\beq{\begin{equation}}
\def\eeq{\end{equation}}
\def\ul{\underline}
\def\ni{\noindent}
\def\nn{\nonumber}
\def\wt{\widetilde}
\def\wh{\widehat}
\def\Tr{\mbox{Tr}\ }

\def\tr{\,\mbox{tr}\,}                  
\def\Tr{\,\mbox{Tr}\,}                  
\def\Res{\,\mbox{Res}\,}                
\renewcommand{\Re}{\,\mbox{Re}\,}       
\renewcommand{\Im}{\,\mbox{Im}\,}       
\def\lap{\Delta}                        

\def\al{\alpha}
\def\be{\beta}
\def\ga{\gamma}
\def\de{\delta}
\def\ep{\varepsilon}
\def\ze{\zeta}
\def\io{\iota}
\def\ka{\kappa}
\def\la{\lambda}
\def\na{\nabla}
\def\pa{\partial}
\def\ro{\varrho}
\def\si{\sigma}
\def\om{\omega}
\def\ph{\varphi}
\def\th{\theta}
\def\te{\vartheta}
\def\up{\upsilon}
\def\Ga{\Gamma}
\def\De{\Delta}
\def\La{\Lambda}
\def\Si{\Sigma}
\def\Om{\Omega}
\def\Te{\Theta}
\def\Th{\Theta}
\def\Up{\Upsilon}


\vspace*{3mm}

\begin{center}

{\Large \bf
Quantum gravity correction, evolution }
\vskip 3mm

{\Large \bf
of scalar field and inflation}

\vskip 8mm

{\sc J. Acacio de Barros$^a$
\footnote{E-mail: acacio@ibitipoca.fisica.ufjf.br},
Nelson Pinto-Neto$^b$
\footnote{E-mail: nen@lca1.drp.cbpf.br},
Ilya L. Shapiro$^{a,c}$
\footnote{E-mail: shapiro@ibitipoca.fisica.ufjf.br} }
\vskip 8mm

{\large\sl
a) Departamento de F\'{\i}sica -- ICE,
Universidade Federal de Juiz de Fora -- MG, Brazil
\vskip 3mm

b) Centro Brasileiro de Pesquisas F\'{\i}sicas, Rua Xavier Sigaud, 150,
CEP-22290-180, Rio de Janeiro, RJ, Brazil
\vskip 3mm

c) Tomsk State Pedagogical 
University, 634041, Tomsk, Russia }

\vskip 8mm
\end{center}

\noindent
{\large \sl Abstract.} $\;\;\;$
We take the first nontrivial coefficient of the Schwinger-DeWitt
expansion as a leading correction to the action of the
second-derivative metric-dilaton gravity. To fix the ambiguities
related with an arbitrary choice of the gauge fixing condition
and the parametrization for the quantum field, one has to use
the classical equations of motion. As a result, the only corrections
are the ones  to the potential of the scalar field. It turns out that
the parameters of the initial classical action may be chosen in
such a way that the potential satisfies most of the
conditions for successful inflation.
\vskip 8mm

\section{Introduction}

The behavior of the scalar field plays a crucial role in most
of the inflationary scenarios. In particular, the special type
of evolution of the
scalar field to a stable vacuum state through the ``slow-roll"
scheme is in the heart of the inflationary scenario. In order to
provide successful inflation, the potential of the scalar
field should satisfy some set of conditions, and it is difficult
to derive such a potential from a reasonable quantum field theory.
According to the modern point of view, inflation takes place
at very high energies where the effects of quantum gravity may
dominate. Unfortunately, a consistent theory of quantum
gravity is unavailable, therefore one has to use some approximate
scheme to explore the important metric-scalar models.

If one supposes that inflation takes place at typical energies below --
but not many orders of magnitude below -- the Planck scale, then it is
possible to use the framework of effective quantum gravity as a
forecast for what happens when quantum gravitational effects are taken
into account. An interesting aspect in the recent development of the
effective approach to quantum gravity \cite{don} (see also \cite{high})
is that the full theory may include many high derivative terms, but they
are not seen at lower energies when the corresponding heavy particles
do not propagate \cite{wei} and the contribution of the corresponding
loops is suppressed \cite{apco}. For this reason, in this work, we
disregard the higher derivative effects and concentrate on the quantum
contributions to the second-derivative local part of the effective
action.

This paper is organized as follows.
In the next section we formulate the action 
and symmetry transformations
for the metric-scalar theory in four-dimensional space-time.  
We choose the theory with the action composed by Einstein-Hilbert 
action with cosmological constant and by the conformal scalar 
field. The introduction of the cosmological constant into the 
classical action is important because, otherwise, one unavoidably
meets the cosmological constant in the vacuum state after the
quantum corrections are taken into account.
As it was discussed in \cite{bek,conf,duco}, this type of 
action possesses an extra symmetry which may be called 
conformal duality. In section 3 we review the previous 
calculation of the first term in the Schwinger-DeWitt expansion
of the one-loop effective action \cite{duco}, discuss its 
gauge-fixing dependence 
and construct the gauge-independent correction to the potential 
for the scalar field. In section 4 the physical conditions for the 
effective potential are formulated and the possibility to have  
successful inflation is explored.
In the last section some conclusions and discussions are 
presented.

\section{ Metric-dilaton actions and its symmetries.}

According to \cite{spec,conf}, all metric-scalar theories can be 
divided into two main classes: with and without local conformal 
symmetry. The models of the first class may be classically equivalent 
to General Relativity (GR) with cosmological constant, while the second
type of models have additional degrees of freedom. All such models 
are related by a reparametrization of the scalar field supplemented 
by a conformal transformation of the metric. Therefore, since the
conformal symmetry is absent and there is no danger of anomaly,
one can safely use any frame for the description of the theory on 
both classical and quantum levels. Let us write the action of the
theory in the form
\beq
S = S_{B(\phi),\lambda} + S_{L(\phi),\tau},    
\label{A1}
\eeq
where $\,B,L\,$ (and later $K$) are some smooth functions of 
$\,\phi$. Both actions $\,S_{B(\phi),\lambda}\,$ and  
$\,S_{L(\phi),\tau}\,$ are classically equivalent to GR.
This means \cite{conf} that
\beq
S_{B(\phi),\lambda} =
\int d^4x \sqrt{-g}\; \left\{\,\,\frac32\,\frac{B_1^2}{B}\,
g^{\mu\nu}\,\partial_{\mu}\phi \,
\partial_{\nu}\phi + B\,R + \lambda\,B^2 \,\,\right\}\, , 
\label{0.1}
\eeq
where $B = B(\phi)$ and $B_1 = \frac{dB}{d\phi}$.
The last action is invariant under the transformation 
\beq
{\bar \phi} =
{\bar \phi} (\phi)
\,,\,\,\,\,\,\,\,\,\,\,\,\,
{\bar g}_{\mu\nu} = 
g_{\mu\nu}\,\frac{B({\bar \phi}(\phi))}{B({\phi})} \, .
\label{c}
\eeq
On the other hand, all such actions (including the case of 
$\,B = (16\pi G)^{-1}=const\, $) may be linked 
with each other by a conformal transformation of the metric
\cite{conf}. Also, the models with a non-constant $B(\phi)$
can be linked by a simple reparametrization of the scalar field.
As a consequence of this, the theory (\ref{A1})
manifests a conformal duality
\beq
S\left[g_{\mu\nu}; B(\phi), K(\phi); \lambda, \tau \right]
=S\left[{\bar g}_{\mu\nu}; L(\phi),
\frac{K(\phi)L(\phi)}{B(\phi)}; \lambda, \tau \right]
\;,\,\,\,\,\,\,\,\,\,\,
{\bar g}_{\mu\nu} = g_{\mu\nu}\,\frac{L(\phi)}{B(\phi)}\; .      
\label{A5}
\eeq
In a particular case one meets the dual symmetry
\cite{bek,conf,duco}:
\beq
{\bar g}_{\mu\nu} \,\leftrightarrow\,
g_{\mu\nu} = {\bar g}_{\mu\nu} \cdot \frac{\ga}{B(\phi)}
\,,\,\,\,\,\,\,\,\,\,\,
\ga\,\leftrightarrow\,\frac{1}{\ga}
\,,\,\,\,\,\,\,\,\,\,\,
B(\phi)\,\leftrightarrow\,\frac{1}{B(\phi)}
\,,\,\,\,\,\,\,\,\,\,\,
\la\,\leftrightarrow\,\tau .
\label{dual}
\eeq
This transformation describes the
inversion of  the coupling constant $\gamma$ and
function $B(\phi)$ and can link the strong and weak coupling
regimes. It should be very interesting to apply this symmetry
to the study of strong-gravity effects, but in the 
present paper we will concentrate on the quantum corrections 
to the classical action (\ref{A1}) and to their influence on
classical cosmological models.


\section{One-loop calculation and mass-shell conditions}

In this section we calculate the one-loop correction to the
effective action for the theory with conformal duality and 
extract its unambiguous part.
Since all the theories with conformal duality (\ref{A1}) differ
from each other by the reparametrization of the scalar field
only, we will perform the calculations for the simplest case 
\beq
S = \frac{1}{\ka^2}\;\int d^4x \sqrt{-g}\; \left\{ \,\frac{3}{2\phi}
\;g^{\mu\nu}\partial_{\mu}\phi\;
\partial_{\nu}\phi + (\phi + \ga)R - V(\phi) \, \right\}\; ,
\label{2.1}
\eeq
where we will put later on $\ga = -1$ and fix
$ - V(\phi) = \la \phi^2 + \tau \ga^2$.

The quantum calculation can be done within the background
field method (see \cite{book} for an introduction).
The features of the metric-dilaton theory require the 
special background gauge, which has been originally
introduced in the similar two-dimensional theory \cite{odsh}
and recently applied for the calculation of the one-loop divergences
in general four-dimensional metric-scalar theory in \cite{spec},
and in the case of the theory (\ref{2.1}) in \cite{duco}.
The starting point is the splitting of the fields
into background $ \, g_{\mu\nu}, \phi \,$ and quantum
ones, $\, {\bar h}_{\mu\nu}, h, \varphi \,$:
\beq
\phi \rightarrow \phi' = \varphi + \ka\,\phi
\;,\;\;\;\;\;\;\;\;\;
g_{\mu\nu} \rightarrow g'_{\mu\nu} + \ka\,h_{\mu\nu}
\;,\;\;\;\;\;\;\;\;\;
h_{\mu\nu} = {\bar h}_{\mu\nu}+\frac{1}{4}\; g_{\mu\nu}h
\;,\;\;\;\;\;\;\;\;\;
h=h_{\mu}^{\mu}\; , \label{2.2}
\eeq
where we divided the quantum metric into the trace and the traceless
parts. The details of the calculations can be found in \cite{duco}. 
Here we just give a brief review of it.
The one-loop correction to the
effective action is given by the standard general expression
\beq
{\bar \Gamma}^{1-loop}=
{i \over 2}\;\Tr\ln{\hat{H}}-i\;\Tr\ln {\hat{H}_{ghost}}
+ {i \over 2}\;\Tr\ln{Y^{\mu\nu}}                       \label{2.3}
\eeq
where $\hat{H}$ is the Hermitian bilinear form of the action 
$S + S_{gf}$ with added gauge fixing term
$$
\left(S + S_{gf}\right)^{(2)}
=\int d^4 x \sqrt{-g}\; \left( {\bar h}_{\mu\nu},\; h, \; \varphi \right)
\;\left( \hat{H} \right)\;
\left( {\bar h}_{\al\be},\; h, \; \varphi \right)^T
$$
where
\beq
S_{gf} = \int d^4 x \sqrt{-g}\;\chi_{\mu}\;Y^{\mu\nu}\;\chi_{\nu}
\label{2.4}\,,
\eeq
$\hat{H}_{ghost}$ is the bilinear form of the action of the 
Faddeev-Popov ghosts. The general form of the
gauge fixing condition and weight function are
\beq
\chi_{\mu} = \nabla_{\la} {\bar h}_{\mu}^{\,\la} +
\beta \,\nabla_{\mu}h + \rho \,\nabla_{\mu} \varphi,
\;\;\;\;\;\;\;\;\;\;\;\;\;\;\;\;\;\;\;\;\;\;\;
Y^{\mu\nu} = - \al \, g^{\mu\nu} \; ,
\label{2.5}
\eeq
where the gauge fixing parameters
$\alpha, \beta, \rho$ are some functions of the background dilaton,
which can be fine tuned to simplify the calculations.
For instance, if one chooses these functions as
\beq
\alpha = - \frac{1}{2}\;\left( \phi + \ga  \right)
,\;\;\;\;\;\;\;\;\;\;\;\;
\beta=-\frac{1}{2}
,\;\;\;\;\;\;\;\;\;\;\;\;
\rho = - \frac{1}{\phi + \ga} \; ,                         \label{2.6}
\eeq
then the bilinear forms $\,{\hat {H}},\,{\hat {H}}_{gh}\,$
can be reduced to minimal operators. For example\footnote{Explicit 
expressions for the matrices in (\ref{2.8})
are given in \cite{duco}.} 
\beq
\hat{H}=\hat{K}\Box+\hat{L}_{\la}\nabla^{\la}+\hat{M} 
\label{2.8}
\eeq
and
\beq
\Tr \ln{\hat {H}} \, = \, \ln \mbox{Det}{\hat {K}}\, + \,
\Tr\ln\left(\hat{1}\Box + {\hat {K}}^{-1} {\hat {L}}^{\mu}\nabla_\mu
+\hat{K}^{-1}\hat{M} \right)\; .
\label{2.9}
\eeq
The first term gives a simple contribution of an ordinary
functional determinant of the $c$-number matrix. 
The second term (just as the bilinear operator of the ghost action
\cite{duco}) has the form of the usual minimal operator
$$
{\hat H}_{min} = {\hat 1}\Box + {\hat E}^\la \na_\la + {\hat D},
$$
to which the Schwinger-DeWitt expansion applies.

Our purpose is to evaluate the quantum gravity corrections to the
classical action 
to the first order in curvature, and in the corresponding second order
in the derivatives of the scalar field. This approximation corresponds
to the first coefficient of the Schwinger-DeWitt expansion.
Thus, for each of the minimal operators we can write
\beq
\ln \mbox{Det} \left(- {\hat H}_{min} \right)\, =
\,-\Tr \,\int_{\epsilon^2}^{\infty}\,\frac{ds}{i\,s}\;
 U_0(x,x';s)\,\sum_{n=0}^{\infty}\;{a}_n(x,x')\,(i\,s)^n \; ,
\label{SdW}
\eeq
where the $\,\Tr\,$ includes taking the usual trace, 
the coincidence limit
$\,x'=x\,$ and covariant integration over $\,x$.
The ultraviolate divergences may be regularized by means of the
dimensional parameter $\ep$. However, since we are considering a
theory which is not fundamental but is supposed to work below some
energy scale, it is natural to introduce the covariant cut-off
directly in the integral (\ref{SdW}). Then, the correction to the
effective action coming from the corresponding operator is nothing
but the functional trace of the $a_1$-coefficient with a factor of
some cut-off parameter $\mu^2$, which is related to $\ep^2$.

In this way, taking into account all the terms in (\ref{2.1}),
we arrive at the following expression
\beq
\Gamma^{(1-loop)}_{1}\, =\,\mu^{2} \int d^4 x\sqrt{-g}\;
\left\{\, {\cal A}(\phi)\;g^{\mu\nu}\;\partial_{\mu}\phi\; 
\partial_{\nu}\phi +
{\cal B}\,(\phi)R + {\cal C}(\phi) \, \right\}\; ,
\label{2.15}
\eeq
where $\mu$ is a dimensional parameter related to
$\epsilon$ in (\ref{SdW}), and
$$
{\cal A}(\phi) =
\frac{1}{( \ga + \phi)^2}\,\left[\,-\frac{1}{4}\,\frac{\ga^2}{\phi^2} 
-6\,\frac{\ga}{\phi} + \frac{23}{2} +\frac{2}{3}\,\frac{\phi}{\ga}
\right] \; ,
$$
\beq
{\cal B}(\phi) =
- \frac{23}{4} +\frac{2}{3}\,\frac{\phi}{\ga}
\,,\,\,\,\,\,\,\,\,\,\,\,\,\,
{\cal C}(\phi) =
\frac{2}{3}\,\la\phi\, \left(\,\frac{5\phi}{\ga} + 1\,\right)
- \frac{2}{3}\,\frac{\phi + 15\ga}{\ga(\ga+\phi)}\,
\left(\,\la\phi^2 + \tau\ga^2\,\right) \; .
\label{res}
\eeq
Eq. (\ref{2.15}), with the coefficients given in eq. (\ref{res}),
is the first term
in the expansion of the one-loop correction to the classical action.
Indeed, this expression does not satisfy the conditions
imposed by conformal duality, and therefore  in this approximation
quantum corrections violate this symmetry. Below we shall analyze
the cosmological implications of the classical action (\ref{2.1})
with the above quantum corrections. 

One has to notice that the expression (\ref{2.15}) is strongly
ambiguous because all the coefficients (\ref{res}) depend on the
choice of the gauge fixing condition, and on the parametrization of
the quantum fields. Thus, before we start to use the above quantum
correction in the cosmological framework, it is necessary to
extract its unambiguous part. The gauge and parametrization dependence
of the effective action is known in general form
\cite{dew-67,vlt} (see also \cite{book}). For us it is enough to know
that this dependence disappears when one uses the one-shell conditions.
Hence, in order to extract some 
invariant quantity from the expression (\ref{2.15}),
one can assume that the background fields $g_{\mu\nu}$ and $\phi$
satisfy classical equations of motion.
As a consequence, the equation for the
scalar field and the trace of the equation for the metric have
the form
\beq
R = 2\ga\tau \,,\,\,\,\,\,\,\,\,\,\,
(\na\phi)^2 - 2\phi\;(\Box \phi) = \phi^2\,S(\phi)\;, 
\label{equa}
\eeq
where
$\,S(\phi) = {4}/{3}\;\left( \tau\,\ga - \la\,\phi \right) \;$
and $(\na\phi)^2=g^{\mu\nu}\partial_\mu\phi\partial_\nu\phi$.
Mathematically, the use of classical equations of motion in the
quantum correction is equivalent to a renormalization of both
metric and scalar field, which preserves the purely classical form
of the kinetic terms for both scalar field and metric. We remark
that keeping
the classical form of the covariant kinetic term for the metric means
removing the quantum correction for the term linear in curvature.
Thus, after the corresponding renormalization of metric and scalar,
the only (unambiguous) quantum correction is the one to the
potential $V(\phi)$.

Technically the problem of going on-shell in the metric-dilaton
theory with a non-minimal interaction is not trivial
(see, for example, \cite{hove,dene}). Here we will follow the
considerations of \cite{duco}.  
For any function $F(\phi)$ and for any solution of the
equations of motion (\ref{equa}) one can write
\beq
\int d^4x \sqrt{-g}\; F(\phi)\;\left\{ {\phi}^{-1}
\left(\na \phi \right)^2 - 2\left(\Box \phi \right) - \phi S(\phi)
\right\} = 0\; .
\label{2.17}
\eeq
Integrating by parts we arrive at
\beq
\int d^4 x \sqrt{-g}\;
\left\{ {\phi}^{-1} F(\phi) + 2 F_1(\phi) \right\}
\left(\na \phi \right)^2
= \int d^4x \sqrt{-g}\; F(\phi)\;\phi S(\phi)\; .
\label{2.18}
\eeq
For any constant $C_1$ the Eq. (\ref{2.18}) is invariant 
under $F(\phi) \rightarrow F(\phi) + C_1/\sqrt{\phi}$. 
Therefore for any solution of (\ref{equa}) the
function $\,\phi^{-1/2}\,$ satisfies the condition
\beq
\int d^4x \sqrt{-g}\; \phi^{-1/2}\;\phi \,S(\phi) = 0\; ,
\label{2.19}
\eeq
and therefore this term can be always disregarded.

In order to have correspondence with (\ref{2.15}), one has to put 
$\,{\phi}^{-1} F(\phi) + 2\, F_1(\phi) = {\cal A}(\phi)$.
The solution for $F(\phi)$ has the form
\beq
F(\phi) =  C_1 \phi^{-1/2} + \phi^{-1/2}\,\int_{\phi_0}^{\phi}
d\phi\;\phi^{1/2}\; {\cal A}(\phi)\; ,
\label{2.21}
\eeq
where the first term can be omitted due to (\ref{2.19}).
Now (\ref{2.21}) must be substituted into (\ref{2.18}) and 
(\ref{2.15}). We then
arrive at the on-shell quantum correction (see \cite{duco}
for further details).
It is important that the corresponding expression does not depend
on the parametrization of the quantum field and on the choice of
the gauge parameters.


Before writing down the 
unambiguous effective potential with the quantum gravity
corrections we make some change of notations:
\beq
\phi=\frac{\chi^2}{12M_{pl}^2}
\,,\,\,\,\,\,\,\,\,\,\,\,\,\,\,\,\,\,\,
\lambda=-\,6f\,M_{pl}^2
\,,\,\,\,\,\,\,\,\,\,\,\,\,\,\,\,\,\,\,
\gamma=1 .
\label{notations}
\eeq
In these new variables,
the kinetic term for the scalar field acquires the common form
$\,1/2(\na\chi)^2\,$.
The effective potential depends on three arbitrary 
parameters: the energy scale 
$\mu$, the scalar coupling $f$, and the
dimensionless cosmological constant $\La$. In
the next section, 
we will check whether there is a region in the space of parameters
$\mu,\La,f$, in which the physical conditions for 
$V_{eff}$ to yield suitable inflation are satisfied. 
Those conditions include, at a first place, 
the boundness of the potential from below. Then the dynamics of the 
scalar field will be governed by the form of the potential, and the 
consistency with the inflationary scenario poses many other 
restrictions. Thus, despite the fact that we have three parameters in hand, 
it is not a trivial matter to satisfy all conditions. The next 
important 
condition on $V_{eff}$ is that in the ground state the value of the 
cosmological constant should be very close to zero. The recent 
estimates show that the observable value of the cosmological constant
density $\rho _{\Lambda} = \Lambda /8\pi G$ is $\rho _{\Lambda} \approx
0.7 \Omega _c \approx (3 {\rm meV})^4$ where $\Omega _c$ is the critical
density of the universe. As we shall see,
this value is many orders smaller than the typical quantities 
characterizing our potential. Thus we can safely use the approximation
in which the value of the cosmological constant is zero in the 
ground state $\chi_0$.
One has to notice that, in the framework of our model, 
it is not possible to fine-tune the value of the cosmological 
constant by adding corresponding ``vacuum'' term. Such a term
should affect the form of the quantum correction in (\ref{2.15}).
Therefore one has to make a more sophisticated fine tuning 
changing the value of $\tau$ (or $\La$) in the starting action.
The local conditions for the potential to have zero ``physical"
cosmological constant in the ground state is
\beq
V_{eff}(\chi_0,\La,f,\mu) = 0\,,\,\,\,\,\,\,\,\,\,\,\,
\frac{d}{d\chi}V_{eff}(\chi_0,\La,f,\mu) = 0,
\label{cosmol}
\eeq 
and these equations for $\La$ must be solved before
one can proceed to the next conditions.

For the subsequent study, it is more adequate to use the 
adimensional variables:
\beq
\tau=\Lambda\cdot M_{pl}^2, 
\,\,\,\,\,\,\,\,\,\,\,\,\,\,\,\,
\chi=x\cdot M_{pl}, 
\,\,\,\,\,\,\,\,\,\,\,\,\,\,\,\,
\mu=y\cdot M_{pl}.
\label{not1}
\eeq
We will explore the values of the coupling $f$ 
ranging from 0 to 1, $y$ is the
cutting scale measured in the units of $M_{Pl}$ 
and $x$ is the dimensionless scalar field also 
measured in the units of $M_{Pl}$. The dimensionless cosmological
constant $\La$ will be found in the next section
from the condition of vanishing of 
effective (physical) cosmological constant.
In terms of $\,x\,$ and $\,y\,$ the effective potential reads
$$
V_{eff} \,=\,\frac{1}{24}\,f x^{4}-\Lambda -
y^2\cdot
\left\{ 
\Lambda \left( 15 - \frac{x^2}{9}\right) -
\frac{f}{12}\,x^2 ( x^2-4) 
-\frac43\,\frac{x^2+180}{x^2+12}\,
\left( \Lambda -\frac{f{x}^{4}}{24}\right) 
\right.
$$
\beq
\left.
+\,\frac89\,
\left(\Lambda +\frac{fx^2}{2}\right) 
\cdot
\left( 1 
+ \frac{115}{72}\,x^2 
+ \frac{17}{2\sqrt{12}}\,x^2 \arctan \left[\frac{x}{\sqrt{12}}\right]
- \frac{199\,x^2}{6(x^2+12)} \right) \right\}\; . 
\nonumber
\eeq

\section{Conditions for slow-roll inflation}

Now we will enumerate the conditions which the potential 
must satisfy in order
to yield a reasonable inflationary scenario \cite{guth,linde,stein}. 
We shall write some of the formulas in terms of the scalar field
$\chi$ from (\ref{notations}) while others will be written in terms 
of $x$.

The slow roll conditions are:
\begin{equation}
\frac{V^{\prime \prime }}{24\pi V}\ll 1,
\label{condition1}
\end{equation}
\begin{equation}
\frac{V^{\prime }}{\sqrt{48\pi }V}\ll 1.  \label{condition2}
\end{equation}
When these conditions are fulfilled, the kinetic term in the 
energy momentum
tensor of the scalar field becomes negligible with respect 
to the potential 
term. The energy density and pressure of the field are given by
\begin{equation}
\rho _{\chi} = \frac{\dot{\chi} ^2}{2} + V,
\end{equation}
\begin{equation}
p_{\chi} = \frac{\dot{\chi} ^2}{2} - V,
\end{equation}
(the gradient terms are damped by
expansion), and under these conditions
we obtain $p_{\chi} \approx -\rho _{\chi}$. 
This corresponds to a vacuum
equation of state, or equivalently, to an effective cosmological 
constant.
The geometry inflates almost exponentially, $a(t) \approx \exp (Ht)$
with $H$ almost constant given by 
\begin{equation}
H^2 \approx \frac{8 \pi V}{M_P}.
\end{equation}
When conditions (\ref{condition1}, \ref{condition2}) cease to be
valid, inflation ends. Hence, they define the initial and final values of
the scalar field, $x_i$ and $x_f$, in whose interval inflation occurs.
In order to solve the flatness and isotropy problems, the geometry must
inflate the necessary amount in order to bring into causal contact
disconnected regions of the early 
Universe which were found to be thermalized
from the observations of the Cosmic Microwave 
Background Radiation (CMBR), 
while making them almost flat. This implies that the number of e-folds 
of growth of the scale factor during this period 
must be greater then 60 \cite{kolb}:
\begin{equation}
-8\pi \int_{x_{i}}^{x_{f}}\frac{V(x)}{V^{^{\prime }}(x)}dx=N>60.
\label{condition3}
\end{equation}
Finally, the magnitude of the density perturbations produced 
in this scenario
at the moment they cross outside the horizon must be 
\begin{equation}
\biggr(\frac{\delta \rho}{\rho}{\biggl)}_{hor} \approx 10^{-5}
\end{equation}
in order to not be in conflict with CMBR
observations. This is because the anisotropies observed by 
the COBE satellite
in the CMBR \cite{smoot} are of this order of magnitude and 
these anisotropies are related
with the density perturbations by means of the Sachs-Wolfe 
effect \cite{sachs} via the equation
\begin{equation}
\frac{\delta T}{T} \approx
\biggr(\frac{\delta \rho}{\rho}{\biggl)}_{hor} ,
\end{equation}
where $\delta T /T$ are the fluctuations in the temperature of the CMBR. 
This is the most difficult condition which inflaton potentials
must satisfy and it reads \cite{kolb}
\begin{equation}
\frac{\delta \rho }{\rho } \sim
-\frac{V^{3/2}(\bar{x})}{M_P^3 V^{\prime }(\bar{x})} \sim 10^{-5},  
\label{condition4l}
\end{equation}
where $\bar{x}$ is the value of the scalar field about $50$ e-folds 
{\em before} the end of inflation, the time where the modes 
of cosmological
interest first cross outside the horizon, which is given by 
\begin{equation}
-8\pi \int_{\bar{x}}^{x_{f}}\frac{V(x)}{V^{\prime }(x)}dx=50.
\end{equation}

We are left with three numeric parameters, $f,$ the coupling strength, 
$\Lambda ,$ 
the cosmological constant, and $y,$ representing the energy
scale, and the question 
is for what parameters we have all above conditions
satisfied for a suitable inflation.

\begin{enumerate}
\item  For $y<0.235$ we can have potentials with the suitable shape and a
local minimum that allows slow rolling.

\item  For $0.245>y>0.235$ we start having other minima and maxima. The
potential does not yield slow-roll inflation.

\item  For $y>0.245$ the quantum term 
dominates for large $x$ and we do not
have stable solutions with a global minimum anymore.

\end{enumerate}

We can now use the values $y<0.235$ in the following way:

\begin{enumerate}
\item  Fix a value for $f;$

\item  Compute the values of $x_{\min }$ and $\Lambda _{\min }$ 
such that $%
V_{eff2},$ the potential, is zero at its global minimum;

\item  Redefine the potential as being $V_{final},$which is the effective
potential $V_{eff2}$ with $\Lambda $ changed by $\Lambda _{\min };$

\end{enumerate}

We now have a potential that has the right form to have 
inflation, and we need
to check whether the conditions are satisfied for different 
values of $f$ and 
$y.$ All conditions are satisfied for $y=0.235$ and $f<10^{-12}.$ 
Some typical plots exemplifying the three possibilities 
are shown in Figures 1 -- 3.

\section{Conclusion}

We have obtained a scalar field potential from the Schwinger-DeWitt
expansion of dilaton gravity which is free of the ambiguities related to
gauge fixing conditions and which can yield successful inflation for
some choices of parameters.  It should be interesting to push further
the model and calculate the power spectra of scalar and density
perturbations, the quadrupole CMBR temperature anisotropies for these
two types of perturbations, and compare with observations. Also,
it should be interesting to see if the other forms of the potential
having more maxima and minima, which, as we have seen, are possible for
other choices of the parameters, can yield inflation but with a $k=-1$
universe.

\vspace{5mm}

\noindent{\large \bf Acknowledgments}
\vskip 1mm

N.P.N. and I.L.Sh. are grateful to the Departamento de Fisica of 
Universidade
Federal de Juiz de Fora for warm hospitality, and also to CNPq  
for financial support. I.L.Sh. work was also supported in part by Russian
Foundation for Basic Research under the project No.96-02-16017.

\newpage

\newpage

\centerline{\large \bf Figure Captions}
\vskip 6mm

\noindent
{\bf Figure 1}. This potential shows a possibility for tunneling.
The values of parameters are: $f = 5\cdot 10^{-13}$ and $y = 0.235$.
The fine-tuned value of $\La$ is $\La = - 0.719\cdot 10^{-12}$ and the 
point of minima of the potential is $x_0=11.43$. 
\vskip 6mm

\noindent
{\bf Figure 2}. This potential fulfills all the
conditions for the slow-roll inflation (without tunelling).
The values for the parameters are: $y = 0.229$ and $ f = 10^{-14}$.
Fine-tuned ``bare" $\La$ is $\Lambda = -0.439\cdot 10^{-15}$ and 
the value of $x$ 
where the potential has a minimum is
$x_0 = 1.423$. Inequality (26) is fulfilled with it being always
less than $0.001$. Inequality (27) is satisfied from $x = 0.0001$ to
$x=0.6$. In this interval the number of e-folds is $99.95$,
satisfying (31).
Equation (35) implies that $\bar{x} = 0.0085$. 
In this case, the left hand side of (34) takes the
approximate value of  $7\cdot 10^{-5}$. Obviously, the desirable 
behavior is provided by the very small value of coupling $f$ 
(which, when reduced, can make the lhs of (34) even smaller)
and consequent smoothness of the potential. Quantum corrections 
are also small and do not change the asymptotic behavior of the
potential at $x\rightarrow\infty$.
\vskip 6mm

\noindent
{\bf Figure 3}. Potential with dominating quantum corrections,
which are not bounded from below. This type of potentials one
always meets for $y$ greater than $0.244$. 

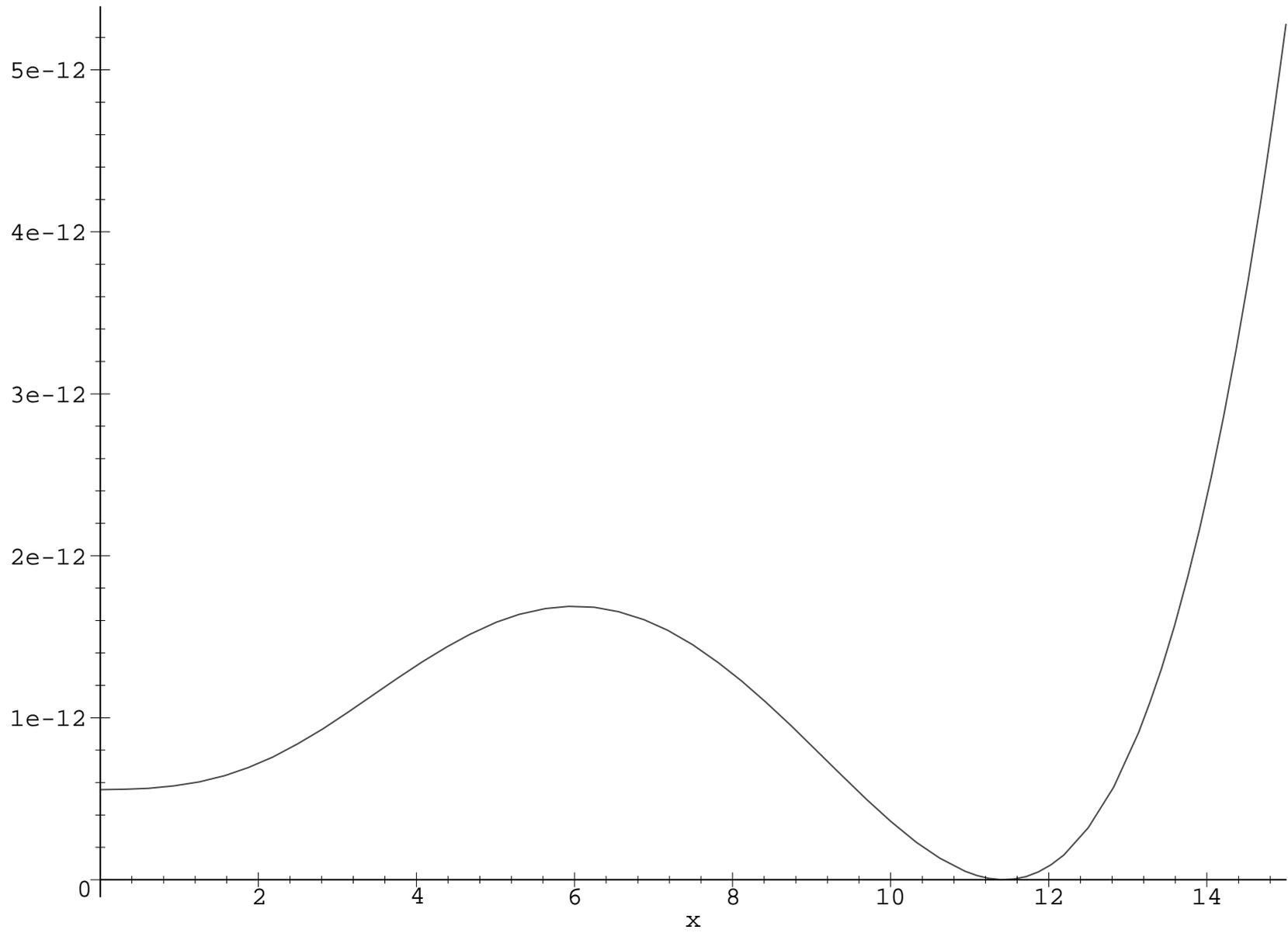
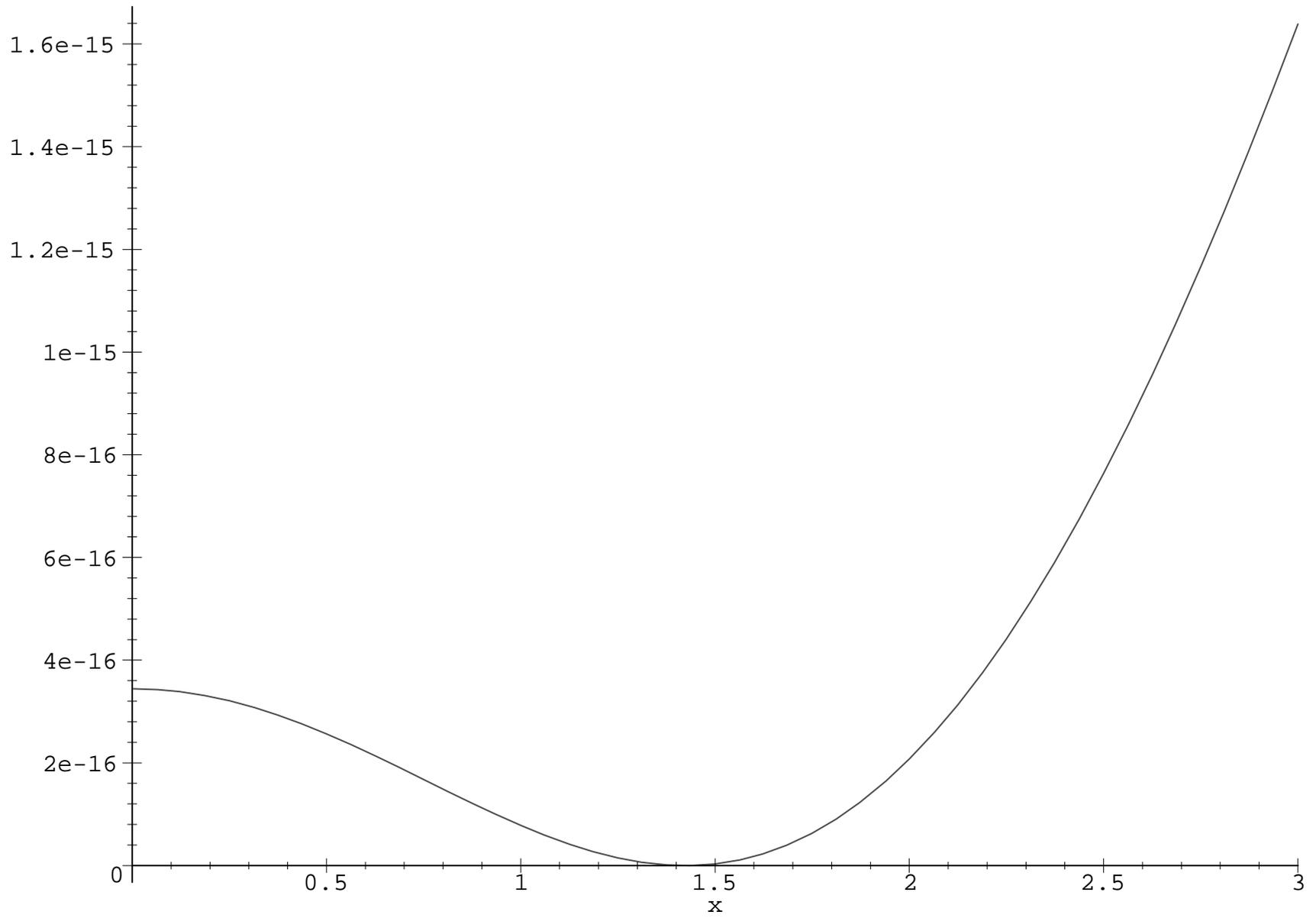
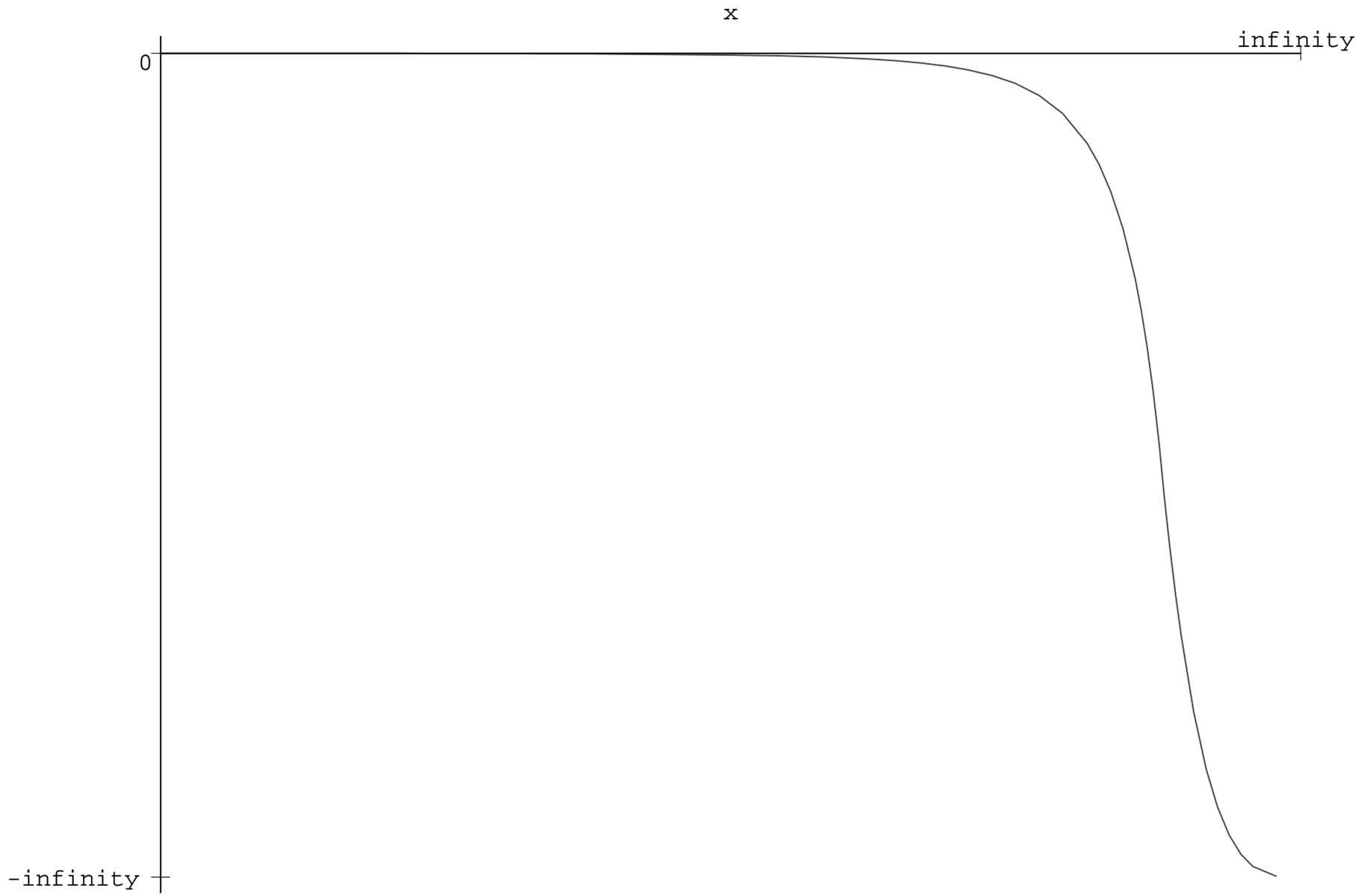
\end{document}